\documentclass[preprint,prb, floatfix, tightenlines, amsmath, amssymb, 
superscriptaddress, longbibliography]{revtex4-1}

\usepackage{graphicx}
\usepackage{color}
\usepackage{dcolumn}
\usepackage{bm}
\usepackage{hyperref}
\usepackage[nolist,nohyperlinks]{acronym}
\usepackage{CJK}
\usepackage{braket}
\usepackage{caption}
\usepackage{ragged2e}
\definecolor{cblue}{RGB}{19,107,192}
\hypersetup{colorlinks=true,linkcolor=cblue,citecolor=cblue,urlcolor=cblue}
\DeclareCaptionLabelSeparator{line}{\,{\bf \textbar}\,}
\DeclareCaptionJustification{fully}{\justify}

\newcommand{\ncnf}{NaCaNi$_2$F$_7$}
\newcommand{\nccf}{NaCaCo$_2$F$_7$}
\newcommand{\ymo}{Y$_2$Mo$_2$O$_7$}

\begin{document}
\title{Continuum of quantum fluctuations in a three-dimensional $S\!=\!1$  
    Heisenberg magnet}
\author{K. W. Plumb}
\affiliation{Institute for Quantum Matter and Department of Physics and 
    Astronomy, The Johns Hopkins University, Baltimore, MD 21218, USA}
\author{Hitesh J. Changlani}
\affiliation{Institute for Quantum Matter and Department of Physics and 
    Astronomy, The Johns Hopkins University, Baltimore, MD 21218, USA}
\author{A. Scheie}
\affiliation{Institute for Quantum Matter and Department of Physics and 
    Astronomy, The Johns Hopkins University, Baltimore, MD 21218, USA}
\author{Shu Zhang}
\affiliation{Institute for Quantum Matter and Department of Physics and 
    Astronomy, The Johns Hopkins University, Baltimore, MD 21218, USA}
\author{J. W. Krizan}
\affiliation{Department of Chemistry, Princeton University, Princeton, NJ
08544}
\author{J. A. Rodriguez-Rivera}
\affiliation{NIST Center for Neutron Research, National Institute of Standards 
    and Technology, Gaithersburg, MD 20899, USA}
\affiliation{Department of Materials Science and Engineering, University of 
    Maryland, College Park, MD 20742, USA}                                             
\author{Yiming Qiu}
\affiliation{NIST Center for Neutron Research, National Institute of Standards 
    and Technology, Gaithersburg, MD 20899, USA}
\author{B. Winn}
\affiliation{NScD Division, Oak Ridge National Laboratory, Oak Ridge, Tennessee 
    37831-6473, USA}
\author{R. J. Cava}
\affiliation{Department of Chemistry, Princeton University, Princeton, NJ
08544}
\author{C. L. Broholm}
\affiliation{Institute for Quantum Matter and Department of Physics and 
    Astronomy, The Johns Hopkins University, Baltimore, MD 21218, USA}
\affiliation{NIST Center for Neutron Research, National Institute of Standards 
    and Technology, Gaithersburg, MD 20899, USA}		
\affiliation{Quantum Condensed Matter Division, Oak Ridge National Laboratory, 
    Oak Ridge, Tennessee 37831-6473, USA}				
\date{\today}
\maketitle
{\bf
Conventional crystalline magnets are characterized by symmetry breaking and 
normal modes of excitation called magnons with quantized angular momentum 
$\hbar$.  Neutron scattering correspondingly features extra magnetic Bragg 
diffraction at low temperatures and dispersive inelastic scattering associated 
with single magnon creation and annihilation. Exceptions are anticipated in 
so-called quantum spin liquids as exemplified by the one-dimensional spin-1/2 
chain which has no magnetic order and where magnons accordingly fractionalize 
into spinons with angular momentum $\hbar/2$. This is spectacularly revealed by 
a continuum of inelastic neutron scattering associated with two-spinon 
processes and the absence of magnetic Bragg diffraction. Here, we report 
evidence for these same key features of a quantum spin liquid in the 
three-dimensional Heisenberg antiferromagnet \ncnf{}.  Through specific heat 
and neutron scattering measurements, Monte Carlo simulations, and analytic 
approximations to the equal time correlations, we show that \ncnf{} is an 
almost ideal realization of the spin-1 antiferromagnetic Heisenberg model on a 
pyrochlore lattice with weak connectivity and frustrated interactions.  
Magnetic Bragg diffraction is absent and 90\% of the spectral weight forms a 
continuum of magnetic scattering not dissimilar to that of the spin-1/2 chain 
but with low energy pinch points indicating \ncnf{} is in a Coulomb phase.  The 
residual entropy and diffuse elastic scattering points to an exotic state of 
matter driven by frustration, quantum fluctuations and weak exchange disorder.}
\newpage
The existence of a spin liquid for isotropically interacting classical spins on 
the pyrochlore lattice was first proposed by Jacques Villain nearly 40 years 
ago.\cite{Villain:1979} Since then, it has been established that the classical 
($S\rightarrow \infty$) Heisenberg antiferromagnet does not undergo any 
magnetic ordering transition.\cite{Harris:1991,Canals:1998, 
    Canals:2000,Moessner:1998_1,Moessner:1998} The magnetic interaction energy 
is minimized by all spin configurations with vanishing magnetization on every 
tetrahedron and the ensemble of these configurations forms a macroscopically 
degenerate, but highly correlated, ground-state manifold. Such a collective 
state is termed a Coulomb phase because coarse-grained spin configurations 
within the manifold form a divergence free vector-field that implies dipolar 
correlations.\cite{ Isakov:2004, Henley:2005, Henley:2010} Experiments probing 
magnetic correlations, and hence the solenoidal field, should include sharp 
pinch point features as in related classical spin ice materials where 
ferromagnetic Ising interactions dominate.\cite{Fennell:2009} Both classical 
spin ice and the classical Heisenberg antiferromagnet may be classified as 
Coulomb phases but, while there is much activity and progress in exploring 
quantum spin ice, much less is understood about the quantum limit of the 
Heisenberg model. There is theoretical evidence that pinch point correlations 
survive, \cite{Harris:1991, Isoda:1998, Canals:1998,Moessner:2006,Huang:2016} 
but the specific character of the ground state and of the magnetic excitations 
are unknown.

The experimental challenge lies in realizing the pyrochlore Heisenberg model in 
a real material. The highly degenerate manifold of the Coulomb phase is 
susceptible to small perturbations \cite{Harris:1991} and lattice instabilities 
\cite{Tchernyshyov:2002} such that at low temperatures the spin liquid phase is 
more often than not supplanted by a broken symmetry phase.  So far, the closest 
realizations of a Heisenberg antiferromagnet on a pyrochlore lattice have been 
found in the cubic-spinels.  Many of these materials exhibit significant 
exchange interactions extending to the second and third nearest 
neighbours.\cite{Conlon:2010}  Magnetic frustration is manifest through 
self-organized independent hexagonal clusters,\cite{Lee:2002,Kamazawa:2004, 
    Chung:2005,Tomiyasu:2008} but a magneto-structural transition severely 
impacts almost half of the magnetic bandwidth.

Extrinsic disorder, in the form of impurity ions, or variations in magnetic 
exchange interactions caused by chemical disorder may also disrupt the spin 
liquid. Generally, these perturbations result in a spin freezing transition at 
low temperatures.\cite{Castella:2001,Saunders:2007,Sen:2015} For example, in 
the Heisenberg pyrochlore \ymo{} weak disorder results in a fully frozen, 
disordered state, with isotropic short range spin 
correlations.\cite{Gardner:1999,Silverstein:2014} Here, we demonstrate that 
disorder is not necessarily fatal to the search for quantum spin liquids and 
can act to only freeze the lowest energy magnetic degrees of freedom.  At 
higher energies a magnetic excitation continuum characteristic of 
fractionalized excitations persists.  

\ncnf{} is one member of a family of recently discovered transition metal 
pyrochlore flourides where charge balance in the neutral chemical structure  
requires an equal mixture of Na$^{1+}$ and 
Ca$^{2+}$.\cite{Krizan:2014,Krizan:2015,Sanders:2016} Diffraction measurements 
probing the average crystal structure indicate that Na$^{1+}$ and Ca$^{2+}$ are 
uniformly and randomly distributed on the A-site of the pyrochlore lattice.  
Magnetic susceptibility measurements reveal Curie-Weiss behaviour, with an 
effective moment of $p_{eff}\!=\!3.6(1)$~$\mu_B$, consistent with $S\!=\!1$, 
and a Curie-Weiss temperature of $\theta_{CW}\!=\!129(1)$~K.\cite{Krizan:2015} 
A spin-glass like freezing transition is observed at $T_f\!=\!3.6$~K in DC and 
AC magnetic susceptibility measurements.\cite{Krizan:2015} This freezing may 
result from the charge disorder that can be expected to generate a random 
variation in the magnetic exchange interactions. For the Heisenberg pyrochlore 
antiferromagnet described by the Hamiltonian $H\!=\!\sum \limits_{ij} 
J_{ij}\mathbf{S}_i\!\cdot\!\mathbf{S}_j$ the freezing temperature provides an 
estimate of the strength of bond disorder $\delta J\!=\!\sqrt{3/8} k_B T_f 
\!=0.19$~meV, for $S\!=\!1$.\cite{Saunders:2007} 

Notwithstanding the glassy features of \ncnf{}, we will provide evidence that a 
quantum spin liquid (QSL) remains a very realistic possibility.
First, the co-existence of a low energy frozen component and the intrinsic 
excitations of a QSL at higher energies is manifest in the magnetic specific 
heat $C_m(T)$. Second, we use theoretical tools including the self consistent 
Gaussian approximation and classical Monte Carlo to perform extensive fits to 
our neutron scattering data and determine the relevant Hamiltonian. We find 
that it is predominantly characterized by a Heisenberg model with small 
additional exchange terms. Third, the presence of a continuum of magnetic 
excitations coupled with the unusually large inelastic spectral weight suggest 
that this $S\!=\!1$ magnet is in the strongly quantum regime. Finally, in the 
absence of a definitive theoretical understanding of the quantum version of the 
pyrochlore Heisenberg antiferromagnet, we explore several scenarios that may be 
consistent with our experimental findings.  

\begin{figure}[htb!]
    \centering
    \includegraphics[]{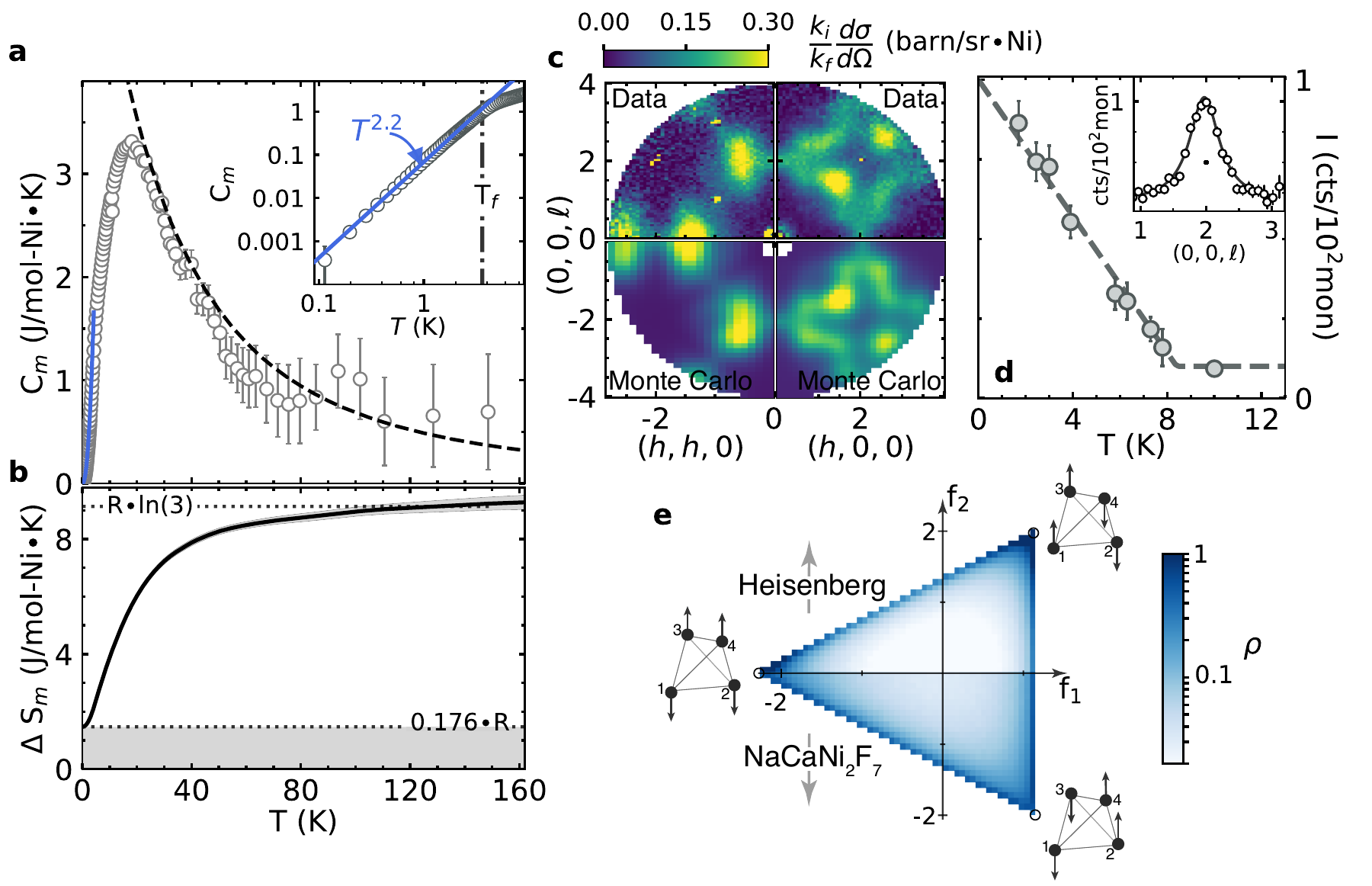}
    \caption{\label{fig:Elastic}{\bf Spin freezing in \ncnf{}}. {\bf a}, 
        Magnetic specific heat. Dashed line is a classical Monte-Carlo 
        simulation. Solid line is a fit to $C_m(T)\!=\!A T^{\alpha}$, with 
        $A\!=\!0.07(1)$ and $\alpha\!=\!2.2(1)$. Inset shows the low 
        temperature region.  {\bf b},  Magnetic entropy obtained by integration 
        of $C/T$ between $T\!=\!150$~K and 100~mK corresponding to  84\% of 
        $R\ln(3)$.  {\bf c}, Diffuse elastic ($E\!=\!0$) magnetic scattering, 
        integrated over the resolution window of $\pm0.37$~meV and obtained by 
        subtracting $T\!=\!40$~K data from that at 1.6~K.  Lower quadrants 
        display disorder and configuration averaged ground state Monte-Carlo 
        structure factors.  {\bf d}, Temperature dependent intensity of the 
        diffuse elastic scattering around $\mathbf{q}\!=\!(0,0,2)$, dashed line 
        is $(1\!-\!T/T_f)^{2\beta}$, with $T_f\!=\!8.2$~K and $\beta\!=\!0.5$.  
        Inset shows the $T\!=\!1.6$~K line shape across the pinch point, 
        integrated over $-0.1\!<\!(h,h,0)\!<\!0.1$, the horizontal dash denotes 
        the instrumental resolution. Error bars in all figures represent one 
        standard deviation.  {\bf e}, Histogram of bond vector order parameter 
        components $(f_1, f_2)$ from classical Monte-Carlo simulations for 
        Heisenberg and exchange model relevant to \ncnf{} including exchange 
        disorder.  Extremal spin configurations corresponding to collinear spin 
        arrangements are shown.}
\end{figure}
Fig.~\ref{fig:Elastic}a shows the magnetic specific heat $C_m(T)$. Beginning 
with the high temperature regime for $T\!>\!18$~K, $C_m(T)$ very closely 
follows the form expected for the classical spin liquid -- Villain's 
cooperative paramagnet --  phase of the Heisenberg antiferromagnet on a 
pyrochlore lattice. Indeed, our classical Monte-Carlo simulation of the 
Heisenberg model, using exchange parameters extracted from analysis of 
inelastic neutron scattering measurements to be discussed below, aligns very 
closely with the data.  In the second regime, where $T$ is of the order of the 
Heisenberg coupling, $C_m(T)$ falls below the classical model and the broad 
maximum at 18~K signals the onset of a collective quantum state.  Finally, a 
third distinct regime is identified below $T_f\!=\!3.6$~K, where a 
discontinuity in the derivative of  $C_m(T)$ occurs.  The approximately 
quadratic power law $T$ dependence below this anomaly is interpreted as a 
consequence of static, or frozen, magnetism below $T_f$.

$C_m\!\propto \! T^2$ for $T\!<\!T_f$ is characteristic of dense frustrated 
magnets where some disorder is present; the exponent appears to be independent 
of the dimensionality of the interacting system.\cite{Ramirez:2000, 
    Nakatsuji:2005, Silverstein:2014} This quadratic temperature dependence 
generally indicates gapless, linearly dispersing modes in two dimensions or 
along nodal lines in momentum space. While the lack of translational symmetry 
implies these do not manifest as coherent modes in neutron scattering 
measurements, the corresponding density of states should be reflected there, 
albeit below the range of energies that we have accessed spectroscopically.  
The low temperature specific heat exponent of $\alpha\!=\!2.2(1)$ could arise 
from the intrinsic low energy sector of a putative QSL. An alternative 
interpretation is attributed to the existence of Halperin Saslow spin waves, 
the normal modes of the frozen state,\cite{Halperin:1977,Podolsky:2009} 
although the presence of line nodes in the dispersion relation is non-trivial.   

In Fig.~\ref{fig:Elastic}b we show the magnetic entropy recovered between 
100~mK and 150~K which saturates at 84\% of the available $R\ln(3)$ for 
$S\!=\!1$.  We interpret the 0.176$R$/spin residual entropy at 100~mK as 
indicating broken ergodicity.  Specifically, we propose that below $T_f$, a 
metastable spin configuration within the Coulomb phase manifold is kinetically 
arrested by the disorder potential so the material no longer explores all 
states of a given energy.  However, most of the magnetic entropy is associated 
with higher energy states. Thus, there is an energy scale above $k_BT_f$ where 
excitations are unaffected by exchange disorder and reflect the site averaged 
spin Hamiltonian of \ncnf{}. This notion is indeed verified through momentum 
and energy resolved neutron scattering measurements, which enable us to
explicitly separate these two components of the spin correlation function. We 
first investigate the frozen component at low energies and then the high energy 
continuum of excitations.  

Figure~\ref{fig:Elastic}c shows the elastic neutron intensity in two 
high-symmetry reciprocal lattice planes of the cubic lattice. The elastic 
magnetic signal is dominated by extended diffuse intensity arising from short 
range correlated spin configurations that are static within the 10~ps time 
window of our measurement. Neutron intensity is concentrated in lobes centered 
on ($2n\pm0.6$, $2n\pm0.6$, 0) positions, where $n$ is an integer. Near $(0 0 
2)$ and $(2 2 0)$, where sharp pinch point features representing long-range 
correlations of the pure Heisenberg model are expected, the momentum 
distribution of the scattering is broader than the experimental resolution.  
The inverse momentum width corresponds to a real-space correlation length of 
$\xi\!=\!6$~\AA{}, or just two nearest neighbour lattice spacings.

Figure~\ref{fig:Elastic}d shows the onset of elastic scattering upon cooling 
below 8~K. This temperature is significantly higher than the 3.6~K $T_f$ 
extracted from susceptibility measurements.\cite{Krizan:2015} Inelastic neutron 
scattering probes the imaginary part of the magnetic susceptibility in the THz 
frequency range, a timescale orders of magnitude faster than AC susceptibility, 
and the upward shift in apparent freezing temperature with the characteristic 
measurement frequency indicates a glass-like transition. We find the momentum 
width of the elastic signal is independent of temperature indicating that 
spatial correlations are unaffected by the freezing transition.  The 
observation of a time-scale dependent $T_f$, temperature independent spatial 
correlations, and residual entropy are consistent with kinetically arrested 
magnetism in \ncnf{}.  Below $T_f$ low energy spin configurations become 
trapped by the disorder potential, resulting in an out-of-equilibrium frozen 
configuration that is a snap-shot of the near degenerate manifold of states.  
Integrating the elastic ($E\!=\!0$) intensity over momentum we find that the 
frozen moment accounts for only $\left| \langle \mathbf{S} 
    \rangle\right|/S\!=\!44\%$ of the saturation magnetization.  Thus,  
magnetism in \ncnf{} at $T\!=\!1.6$~K is predominantly dynamic. Such a small 
fraction of frozen magnetization is comparable to two-dimensional frustrated 
magnets\cite{Lee:1997} but in three-dimensional magnets is unique to \ncnf{}.

To better understand the nature of the frozen low temperature state, we have 
carried out classical Monte-Carlo simulations of the Heisenberg Hamiltonian 
relevant to \ncnf{}.  Random bond disorder was included by sampling from a box 
distribution, with a half width of $\delta J\!=\!0.19$~meV and exchange 
parameters extracted from an independent analysis of inelastic neutron 
scattering data. In figure~\ref{fig:Elastic}c we compare the measured elastic 
scattering with the corresponding numerically modeled signal.  The high 
fidelity fit gives confidence in our optimized magnetic Hamiltonian. To gain 
additional insight we complement these results with a study of local metrics 
for individual tetrahedra on the pyrochlore lattice.

In the absence of disorder, the energy of the classical Heisenberg Hamiltonian 
is minimized by all states with zero total spin per tetrahedra, 
$\mathbf{S}_{\mathrm{tot}}\!= \!\sum \limits_{i=1}^4 \mathbf{S}_i\!=\!0$. We 
find the lowest energy states for the bond-disordered Heisenberg Hamiltonian 
with small anisotropic exchanges relevant to \ncnf{} also fall within the  
$\mathbf{S}_{\mathrm{tot}}\!=\!0$ manifold [see supplemental information].  
This manifold is parameterized by the order parameters  $f_1 = 
\left[\left(\mathbf{S}_1 +\mathbf{S}_2\right) \!  \cdot \left(\mathbf{S}_3 
        +\mathbf{S}_4\right)\!-\!2\mathbf{S} _1 \!  \cdot \!  \mathbf{S}_2 \!  
    - \!2\mathbf{S}_3\!\cdot\!\mathbf{S}_4\right]/\sqrt{12}$ and $f_2\!=\!  
\left(\mathbf{S}_1\!\cdot\!\mathbf{S}_3\! + \! \mathbf{S}_2\!  \cdot \!  
    \mathbf{S}_4\! - \! \mathbf{S}_2\!\cdot\!\mathbf{S}_3\! - \!  
    \mathbf{S}_1\!  \cdot \!  \mathbf{S}_4  \right)/2$.  
\cite{Tchernyshyov:2002, Chern:2008}  The statistical distribution of $f_1$ and 
$f_2$ over a Monte Carlo ensemble of tetrahedra provides a local 
characterization of the particular $\mathbf{S}_{\mathrm{tot}}\!=\!0$ spin 
configuration. Such histograms of $(f_1, f_2)$ extracted from our Monte-Carlo 
simulations are shown in figure~\ref{fig:Elastic}e where possible values span 
an equilateral triangle in the $(f_1, f_2)$ plane. Tetrahedra with pairs of 
antiparallel spins lie along the triangular edges while collinear spin 
configurations are at the vertices.  The classical Heisenberg (only) model with 
weak bond disorder is glassy \cite{Saunders:2007} with a tendency to form 
locally collinear states; this is confirmed by the results in the top half of 
Fig.~\ref{fig:Elastic}e.\cite{Castella:2001}  The enhanced density along the 
boundaries, and away from the corners, of the lower part of the triangle in 
Fig.\ref{fig:Elastic}e indicates the tendency to form configurations of 
pairwise collinear spins when additional small anisotropic interactions 
specific to \ncnf{} are added.

\begin{figure}[]
    \includegraphics[]{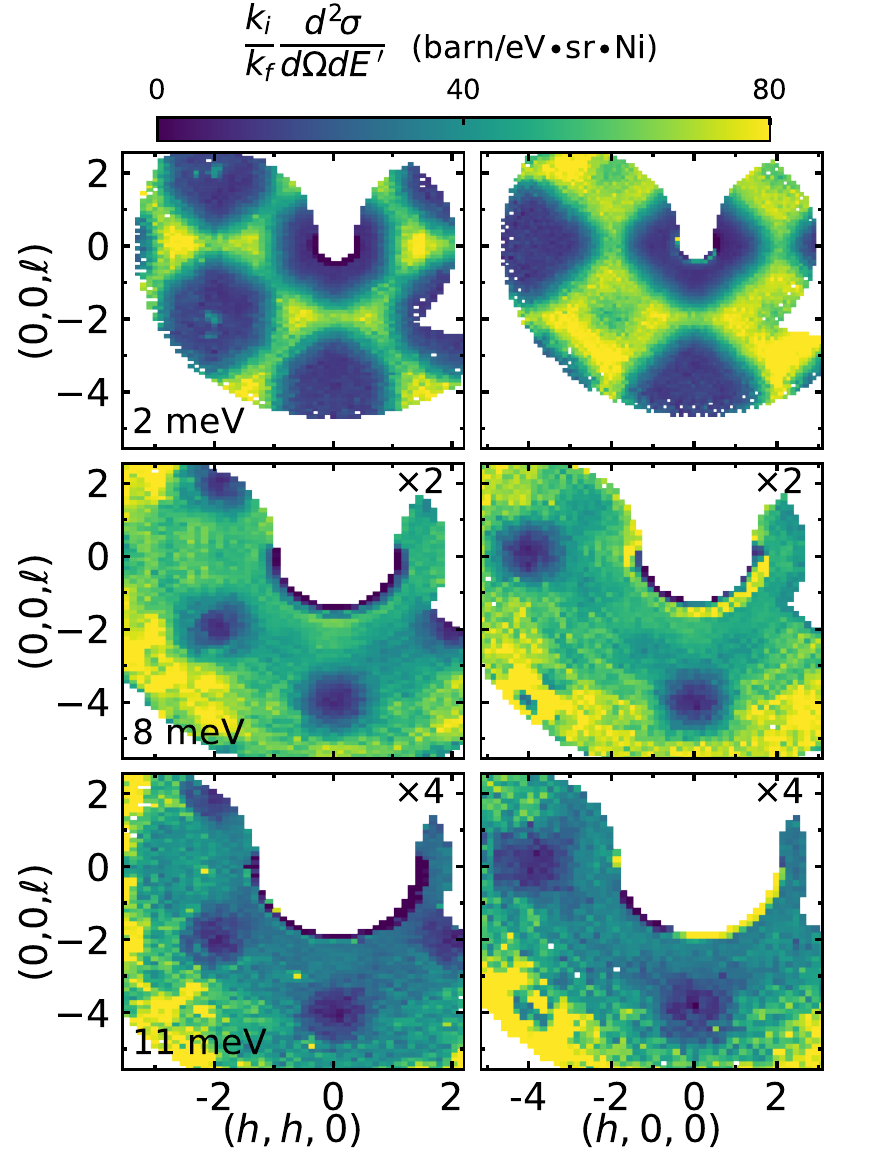}
    \caption{\label{fig:Inel_slices}{\bf Separation of magnetic energy scales 
            in \ncnf{}.} Momentum and energy dependence of inelastic magnetic 
        scattering in \ncnf{} for the $(h,h,\ell)$ and $(h,k,0)$ scattering 
        planes at $T\!=\!1.5$~K.  Each slice was integrated over an energy 
        transfer range of $\pm$0.25 meV.  Above energy transfers of 0.5~meV the 
        dynamic magnetic correlations form a ``bow tie'' pattern in momentum 
        space. The sharp pinch point like features around $(2,0,0)$ and 
        $(2,2,0)$ positions indicate that the net magnetisation per tetrahedron 
        vanish in the Coulomb phase.  Above energies of 5~meV the scattering 
        forms a broad continuum with no intensity around the $\Gamma$ points.}
\end{figure}
In Fig.~\ref{fig:Inel_slices} we present the momentum and energy dependence of 
inelastic magnetic scattering for \ncnf{}. In contrast to the distinct maxima 
in the elastic scattering (Fig.~\ref{fig:Elastic}), the dynamic structure 
factor forms a bow tie pattern with pinch points characteristic of dipolar spin 
correlations.  The scattering closely resembles expectations for the Heisenberg 
antiferromagnet on the pyrochlore lattice \cite{Isakov:2004, Henley:2005, 
    Conlon:2010} but with important deviations, including a slight momentum 
broadening and reduction of intensity around the pinch points. Magnetic 
scattering evolves into a continuum with a well-defined momentum structure at 
higher energies.  The highest energy magnetic excitations are spread everywhere 
in momentum space except at the $\Gamma$ point where neutron intensity is 
precluded for a Heisenberg model.

\begin{figure}[]
        \includegraphics[]{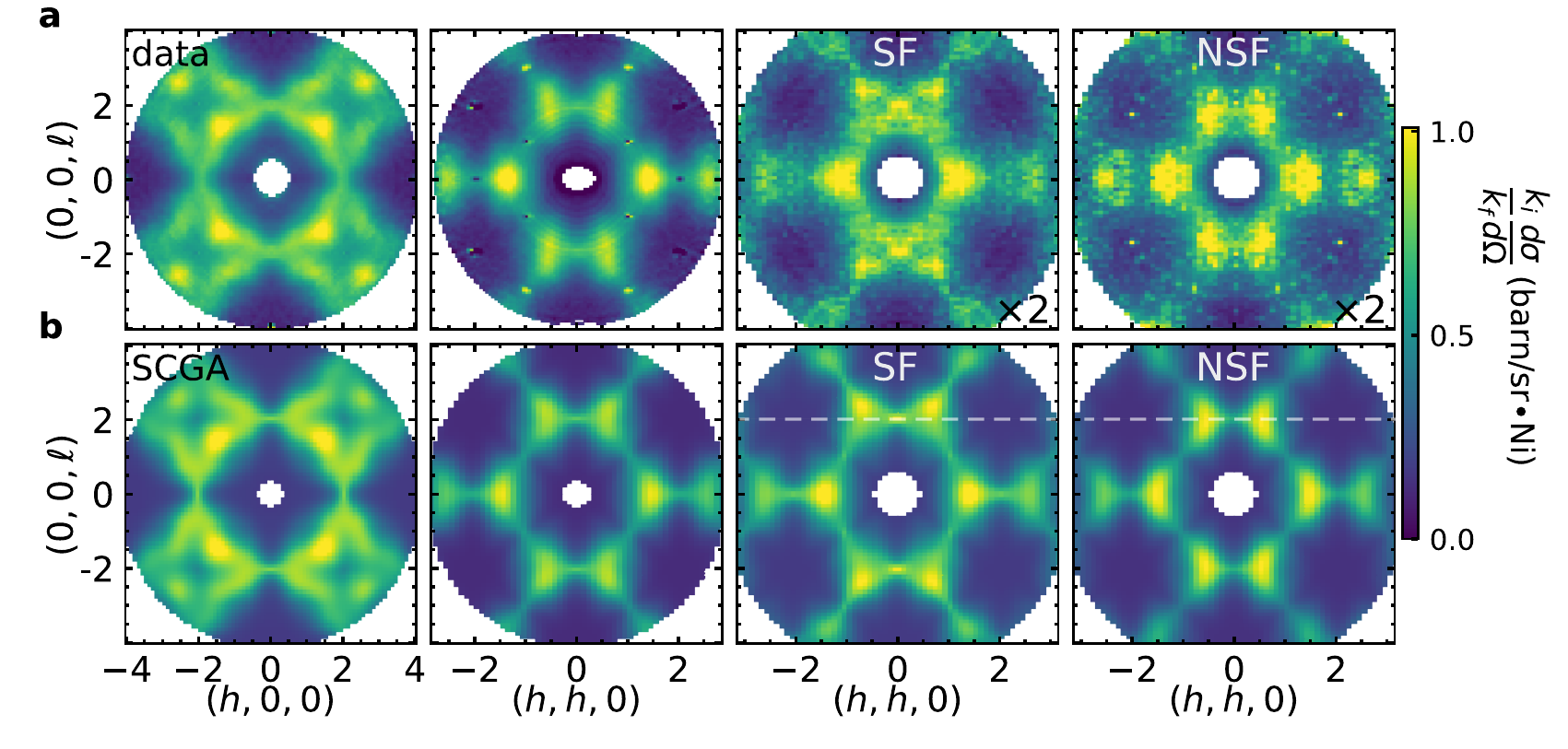}
        \caption{\label{fig:StaticSF}{\bf Equal time structure factor in 
                \ncnf{}}.  {\bf a}, Measured neutron cross-section integrated 
            over the range $0\!<\!E\!<\!14$~meV.  Polarized neutron 
            measurements are labelled by SF, which measures components of the 
            dynamic spin correlation function that are perpendicular to the 
            $(h,h,\ell)$ scattering plane, and NSF, which measures the 
            component of the dynamics spin correlation function polarized 
            within the $(h,h,\ell)$ scattering plane and perpendicular to 
            momentum transfer. {\bf b}, Energy integrated neutron cross-section 
            calculated using the self-consistent Gaussian approximation (SCGA) 
            and exchange parameters $J_{1}\!=\!J_{2}\!=\!3.2(1)$~meV, 
            $J_{3}\!=\!0.019(3)$~meV  $J_{4}\!=\!-0.070(4)$~meV, $J_{NNN}\!=\!  
            -0.025(5)$~meV. Dashed lines delineate plane of asymmetry in the SF 
            scattering. The dipole approximation for the Ni$^{2+}$ magnetic 
            form factor\cite{Brown:06} was used when converting the calculated 
            $\cal{S}(\mathbf{q})$ to a neutron cross-section.}
\end{figure}

Fig.~\ref{fig:StaticSF}a shows the equal time structure factor $S(\mathbf{q})$ 
obtained from the energy integrated magnetic neutron scattering intensity.  
More detailed information is provided by polarized neutron scattering in the 
$(h,h,\ell)$ plane which is sensitive to spin components within the 
$(h,h,\ell)$ reciprocal lattice plane for the non-spin-flip (NSF) channel, and 
along $(1,-1,0)$ for the spin-flip channel (SF). The similarity of SF and NSF 
magnetic neutron intensities in figure~\ref{fig:StaticSF}a is evidence of a 
near spin-space isotropic manifold and immediately rules out single 
ion-anisotropy terms.  Weakly anisotropic interactions are revealed by two 
features of the polarized intensity. First, the SF scattering exhibits a 
pronounced asymmetry of the lobes of intensity centered on $(\pm0.6,\pm0.5,2)$ 
positions about the dashed line parallel to $(1,1,0)$ and passing though 
$(0,0,2)$ indicated in figure~\ref{fig:StaticSF}b.  Second, the NSF intensity 
is diminished around the $(0,0,2)$ pinch point positions.

We have analyzed the energy integrated neutron spectra using a self-consistent 
Gaussian approximation (SCGA) for the equal time structure 
factor\cite{Conlon:2010} using the full symmetry allowed nearest-neighbour 
Hamiltonian $H\!=\!1/2\sum\limits_{ij}J^{\mu\nu}_{ij}S^{\mu}_iS^{\nu}_j$, where 
the 3$\times$3 interaction matrix $J^{\mu\nu}$ is parameterized by four 
independent terms: $J_1$, $J_2$, $J_3$, and $J_4$,\cite{Ross:2011} in 
additional to next nearest neighbour Heisenberg exchange $J_{NNN}$. A symmetry 
allowed biquadratic exchange term was not included in our analysis. We find the 
best global fit of the measured equal time factor with the SCGA using the 
exchange parameters: $J_{1}\!=\!J_{2}\!=\!3.2(1)$~meV, $J_{3}\!=\!0.019(3)$~meV  
$J_{4}\!=\!-0.070(4)$~meV, and  $J_{NNN}\!=\!  -0.025(5)$~meV. This set of 
parameters yields a Curie Weiss temperature $\tilde{\Theta}_{\rm 
    CW}\!=\!-150$~K which may be compared with the experimentally determined 
value of $\Theta_{\rm CW}\!=\!-129(1)$~K.\cite{Krizan:2015} Details of the 
fitting procedure are contained in the supplementary information and the 
resulting modeled neutron intensity is shown in figure~\ref{fig:StaticSF}b.  
Although the SCGA is an approximate procedure, we find exceptional agreement 
between the model and data.  Furthermore, these exchange parameters were 
directly input into the classical Monte-Carlo simulations which builds further 
confidence in the SCGA.  Thus, the spin Hamiltonian for \ncnf{} very closely 
approximates the $S\!=\!1$ Heisenberg antiferromagnet on the pyrochlore 
lattice, perturbed by small symmetric and antisymmetric exchange anisotropies 
as well as next nearest neighbour interactions. 

A number of theoretical investigations have shown that small perturbations in 
the classical Heisenberg model, $J^{\prime}$, in the form of exchange 
anisotropies or further neighbour interactions,  result in a magnetically 
ordered phase below temperatures of the order 
$J^{\prime}S^2$.\cite{Harris:1991, Reimers:1991, Elhajal:2005, Chern:2008, 
    Conlon:2010} In \ncnf{} these perturbations are significantly smaller than 
the freezing temperature, such that any lower temperature transition is 
pre-empted by spin freezing and inaccessible to experiment.  Indeed, our 
classical Monte-Carlo simulations for the anisotropic Hamiltonian relevant to 
\ncnf{}, but in the absence of exchange disorder, do not indicate long-range 
ordering above $T\!=\!500$~mK.  This temperature is well below the broad 
maximum in specific heat where quantum mechanical fluctuations become important 
and our classical simulations are no longer strictly valid.

\begin{figure*}[htb!]
    \includegraphics[]{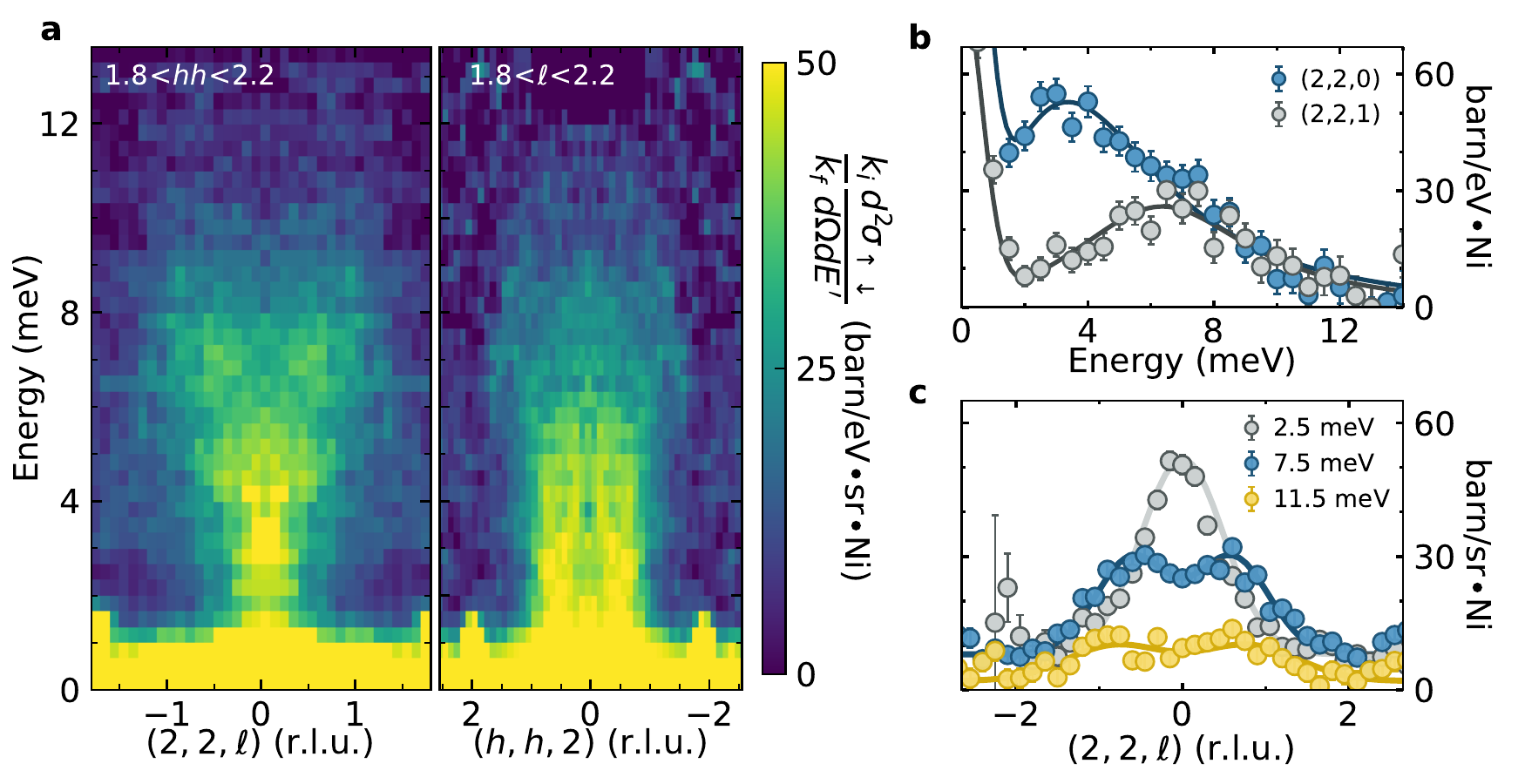}
     \caption{\label{fig:spectra}{\bf Magnetic excitations in \ncnf{}.} {\bf 
             a}, Energy-momentum cuts through the spin-flip portion of the 
         polarized neutron scattering cross-section at $T\!=\!1.6$~K. {\bf b}, 
         Constant momentum cuts of the spin-flip cross-section through a pinch 
         point at $\mathbf{q}\!=\!(2,2,0)$ and nodal point at (2,2,1) 
         integrated over $\ell\!\pm\!0.2$. Solid lines are a fit to the sum of 
         a Lorentzian function centered on the elastic line and a damped 
         oscillator form $S\left(E \right)\!=\!\frac{(n+1)2\Gamma E}{\left(E^2 
                 - E_q^2\right)^2 + (2\Gamma E)^2}$ where $n$ is the thermal 
         population factor, $\Gamma$ a relaxation rate, and $E_q$ the 
         characteristic energy scale. {\bf c}, Constant energy transfer cuts, 
         integrated over $E\!\pm\!0.25$~meV, showing the energy evolution of 
         momentum dependent scattering which bifurcates above 5~meV and 
         precludes a simple factorization of the dynamic structure factor as 
         $S({\bf q},\omega)\!=\!S({\bf q})f(\omega)$.}
\end{figure*}

In figure~\ref{fig:spectra} we present the momentum and energy resolved 
spin-flip neutron scattering cross-section. For our experiment this 
cross-section is sensitive to magnetic scattering and nuclear incoherent 
scattering, thus data in Fig.~\ref{fig:spectra} are representative of the 
dynamic structure factor uncontaminated by coherent non-magnetic scattering.  
The magnetic excitations form a continuum that extends over an energy bandwidth 
of $\sim$12.5~meV$\sim\!4J_1S$.  Along the $(h,h,2)$ direction, transverse to 
the pinch points,\cite{Henley:2005} the inelastic neutron intensity is 
relatively featureless.  However, along the (2,2,$\ell$) direction, 
longitudinal to the pinch points, the magnetic scattering is more structured 
and, importantly, does not factorize as found theoretically for the classical 
limit of the Heisenberg model on the pyrochlore lattice. \cite{Conlon:2009} In 
the constant momentum and energy transfer cuts plotted in  
Fig.~\ref{fig:spectra} b and c very broad dispersive ridges are observed that 
are reminiscent of heavily damped spin-waves. While the spectrum is gapless 
down to the 0.17~meV scale set by our finest energy resolution measurements, 
the dynamic structure factor is peaked at finite energy transfers and can be 
fit with the spectral form of an over-damped harmonic oscillator.  The 
characteristic energy scale disperses from $E_q\!=\!4.8$~meV$\sim J$ at the 
pinch point ${\bf q}\!=\!(2,2,0)$, to $E_q\!=\!7.8$~meV at the nodal point 
${\bf q}\!=\!(2,2,1)$.  This spectrum distinguishes \ncnf{} from recent 
theoretical treatments of the semi-classical Heisenberg model which find a 
purely diffusive response at the pinch points \cite{Huang:2016}. The absence of 
inelastic scattering at the $\Gamma$ point and our polarized neutron 
measurements rule out any sizable single-ion anisotropies that could explain 
the peak in spectral weight at non-zero energy transfers. The only energy scale 
large enough to account for the resonance is the exchange interaction $J_1$. 

Disorder in \ncnf{} is small such that its effect is only to rearrange the low 
energy part of the spectrum for $E\!<\!k_BT_f$ and the underlying translational 
invariance of the Heisenberg spin Hamiltonian can be expected to prevail.  
Indeed, we find that \ncnf{} forms a Coulomb-like phase, with $\mathbf{S}_{\rm 
    tot}\!\approx\!0$ for every tetrahedron. The high energy excitations 
correspond to propagating defects that violate this condition, and our 
observation of an excitation continuum means that such defects in the 
Heisenberg model cannot propagate as coherent quasiparticles carrying an 
angular momentum of $\hbar$. 

A conservative interpretation is that the dispersing modes are overdamped spin 
waves of an underlying classical magnetic order, disrupted in \ncnf{} by 
exchange disorder. Since the frozen spin configurations feature non-collinear 
interacting spins, single particle $S\!=\!1$ magnon excitations can decay from 
interactions with multimagnon states to form a continuum of 
scattering.\cite{Zhitomirsky:2013} Such a scenario may be appropriate for the 
related pyrochlore XY antiferromagnet \nccf{}.\cite{Krizan:2014} Elastic 
magnetic neutron scattering from  \nccf{} resembles that of an ordered 
antiferromagnet, consistent with the non-collinear magnetic structure favoured 
by an order-by-disorder mechanism.\cite{Ross:2016} This order develops at a 
temperature coincident with a broad peak in the magnetic specific heat, which 
constitutes the total magnetic entropy of the $J\!=\!1/2$ magnetic moments 
formed by Co ions.  Furthermore, the fraction of elastic magnetic neutron 
intensity is 0.3(1),\cite{Ross:2016} almost exactly as expected for an ordered 
$S\!=\!1/2$ magnet ($S^2/S(S+1)\!=\!1/3$). Thus, it appears that in \nccf{} 
exchange disorder truncates the magnetic correlations of the classical 
antiferromagnetic order favoured by the underling Hamiltonian.  This is 
distinct from the Heisenberg Hamiltonian we infer for \ncnf{}. It does not 
favour a magnetically ordered state, consistent with the magnetic specific heat 
and the strong inelastic magnetic neutron scattering.  

In contrast to its Co counterpart, \ncnf{} does not show a full recovery of the 
fraction of elastic magnetic neutron intensity; rather, the significant 
proportion of inelastic spectral weight suggests a quantum fluctuating state.  
For a QSL the magnetic spectral weight at $T\!=\!0$ must be entirely accounted 
for by the excitations and there can be no truly elastic scattering. For a 
semi-classical state the elastic scattering should carry a fraction of 
$S^2/S(S+1)$ of the spectral weight, which for $S=1$ equals $1/2$.  By 
integrating the measured dynamic spin correlation function 
$S\left(\mathbf{q},E\right)$ over momentum and energy, including the elastic 
diffuse magnetic scattering,  we recover the total spectral weight of $3\!\int 
{\cal S}\left(\mathbf{q},E\right) dEd^3q\!=\!13(1)$, which is consistent with 
the $(3.7)^2\!=\!13.7~\mu_B^2$ effective moment extracted from the magnetic 
susceptibility.~\cite{Krizan:2015} Comparing the spectral weight for elastic, 
$E\!<\!0.7$~meV, and inelastic scattering, $0.7\!<\!E\!<\!14$~meV, we find 
$\sim\!90\%$ of the magnetic scattering is inelastic in the low $T$ limit.  
This significantly exceeds the 50\% mark for a semi-classical ground state and 
is direct evidence of a spin system dominated by quantum fluctuations. 

Our experimental and theoretical results admit the possibility that \ncnf{} is 
a QSL driven to freezing by weak exchange disorder.  At energies above 
$E\!=\!k_bT_f$ the continuous spectrum indicates the absence of coherent 
quasiparticles carrying angular momentum $\hbar$ and is consistent with the 
fractionalization of a spin flip excitation into weakly interacting 
quasiparticles with angular momentum $\hbar/2$. The fact that the residual 
entropy as a fraction of the total spin entropy ($\Delta 
S/R\ln(3)\!=\!16(4)\%$) is within error bars of the fraction of the total 
spectral weight contained in elastic scattering ($\langle m_{elastic} 
\rangle^2/g^2S(S+1))\!=\!10(2)\%$) indicates the exchange disorder associated 
with the mixed Na/Ca site induces a non-ergodic low energy landscape for these 
quasiparticles. Such a separation of energy scales between frozen and 
fluctuating components is observed in other materials that support QSLs.  For 
example in the one-dimensional $S\!=\!1/2$ chain KCuF$_3$, the spinon continuum 
is observable at high energies even in the N\'eel ordered 
state.\cite{Lake:2005}

A QSL remains a realistic contender for the ground state of \ncnf{}, but 
preciously little is known theoretically about the $S\!=\!1$ Heisenberg model 
on the pyrochlore lattice. Our finding that, at the classical level, the frozen 
state involves tetrahedra with quasi-static anti-parallel pairs of spins at low 
temperatures points to a quantum scenario where these pairs correspond to a 
singlet covering of the pyrochlore lattice. Since there are exponentially many 
such coverings, effects analogous to those studied extensively for dimer models 
may play a role in explaining the specific value of the residual 
entropy.~\cite{Moessner:2006} In addition, the $S\!=\!1$ Heisenberg model 
admits other rich possibilities. One such picture that might be pursued 
involves fluctuating Haldane/AKLT loops decorating the pyrochlore 
lattice.~\cite{Wang:2015} In the AKLT construction, each $S\!=\!1$ degree of 
freedom is built from two $S\!=\!1/2$ objects and the collective quantum state 
is projected to the $S\!=\!1$ subspace.  Loops are constructed by joining 
neighbouring spins into singlets across each bond.  A single spin flip 
excitation will break this bond, fracturing the loop and leaving a chain with 
two free $S\!=\!1/2$, one at each end.  These end states may then act as bulk 
fractionalized excitations that are deconfined within the quantum superposition 
of fluctuating loop coverings.\cite{Savary:2015} In the absence of any 
significant lattice distortion or cluster formation, such liquid like states 
remain a realistic possibility on the pyrochlore lattice and could help to 
understand our observation of residual entropy and continuum scattering in 
\ncnf{}.

\noindent
\section*{Methods}
\noindent
The identification of any commercial product or trade name does not imply 
endorsement or recommendation by the National Institute of Standards and 
Technology.

\textbf{Specific Heat}. Heat capacity measurements were conducted using a 
Quantum Design PPMS with a dilution insert for temperatures between 100 mK and 
4 K, and standard insert for temperatures between 2 K and 270 K.   All 
measurements were carried out on the same 5~mg single crystal using the 
adiabatic pulse method.  The non-magnetic contribution was determined by 
scaling the measured specific heat of the iso-structural compound 
NaCaZn$_2$F$_7$ by the relative Debye temperatures. 

\vspace{2mm}
\noindent\textbf{Neutron Scattering.} All neutron scattering measurements were 
performed on the same 3~g single crystal, grown as described 
elsewhere.\cite{Krizan:2015} Unpolarized neutron scattering measurements were 
preformed on the MACS spectrometer\cite{MACS} at the NIST Center for Neutron 
Research.  The neutron momentum transfer is indexed using the Miller indices of 
the cubic unit cell, $(h,k,\ell)\!=\!(2\pi/a,2\pi/a,2\pi/a)$, where 
$a\!=\!10.31$~\AA.  Measurements were conducted with the sample oriented in 
both the $(h,h,\ell)$ and $(h,k,0)$ scattering planes.  Elastic ($E\!=\!0$) 
measurements were conducted with the monochromator in a vertical focusing 
configuration using a neutron energy of 5~meV. Two configurations were utilized 
for inelastic measurements, both with the monochromator in double focusing 
mode. For energy transfers below 1.4~meV, MACS was operated with fixed final 
energy of 3.7~meV and Be BeO filters before and after the sample respectively.  
For energy transfers above 1.4 meV we used a fixed final energy of 5~meV with a 
Be filter after the sample and no incident beam filter. The data for energy 
transfers above 1.4 meV was corrected for contamination from high-order 
harmonics in the incident beam neutron monitor.  

Polarized neutron scattering measurements were carried out on the HYSPEC 
spectrometer\cite{HYSPEC} at the Spallation neutron Source at Oak Ridge 
National Lab.  An incident neutron energy of 17~meV was selected using a Fermi 
chopper rotating at 240 Hz resulting in an energy resolution of $\delta 
E\!=\!1.4$~meV on the elastic line.  The incident neutron beam polarization was 
defined using a vertically focusing Heusler monochromator while the outgoing 
beam polarization was selected using a radially collimating supermirror array.  
All polarized measurements were conducted with the guide field applied 
perpendicular to the $(h,h,l)$ scattering plane, along the $(1,-1,0)$ 
direction.  In this configuration, spin-flip scattering measures the component 
of magnetic cross section that is polarized within the scattering plane, while 
non-spin-flip measures the out-of-plane component. The flipping ratio measured 
on a $(4,4,0)$ nuclear Bragg peak was 16.  All data reduction and analysis was 
carried out using the Mantid software suite.\cite{Arnold:2014}

Measured neutron count rates from both instruments were placed into absolute 
units of the neutron scattering cross-section using incoherent elastic 
scattering from the sample. The scale factor for conversion to absolute units 
was additionally cross-checked against measurements from a Vanadium standard. 

\vspace{2mm}
\noindent\textbf{Numerical Methods.} We fit the static structure factor from 
the neutron scattering data to the corresponding prediction of the 
self-consistent Gaussian approximation~(SCGA) at 1.6 K, to obtain the parameter 
set in the main text. Details of the method, including the cost function and 
error analysis are discussed in the supplementary information. The results of 
the SCGA are complemented by classical Monte-Carlo~(MC) calculations, for both 
the specific heat and the structure factor. MC simulations used single spin 
updates for continuous spin on pyrochlore lattices (with 16 site cubic unit 
cells) of size $N=16 L^3$ for $L=3$ to $L=10$. For determining the classical 
ground state of the fitted spin Hamiltonian, parallel tempering 
MC~\cite{Nemoto:1996} was carried out with $T_{min}=0.01$ K and $T_{max}=1$ K 
with the number of replicas $N_r=\sqrt{N} \ln \Big( T_{max}/T_{min}\Big)$ 
(approximately 100 for $L=3$ and 400 for $L=8$, the two sizes studied 
extensively, see supplementary for more analyses) and the simulation carried 
out for $10^8$ total steps. With the lowest energy configurations encountered 
in this finite Monte-Carlo run, further iterative minimization was performed to 
accelerate the approach to the classical ground state. For these optimized spin 
configurations (many of which are local minima) two-component local order 
parameters ($f_1$ and $f_2$) are calculated on all $N/4$ tetrahedra of a fixed 
orientation ("up"). This is repeated for 50 -- 100 bond-disorder realizations 
and the combined data set, including all tetrahedra and disorder realizations, 
is used to obtain the 2D histogram in Fig.~\ref{fig:Elastic}. The static 
structure factor from these low energy zero temperature configurations, for 
$L=8$, was averaged to obtain an estimate of the elastic cross-section. Further 
details of all methods and algorithms employed are discussed in the 
supplementary information. 
    
\section*{Acknowledgments}
We are grateful to Yuan Wan for enlightening discussions. This work benefited
from many insightfull comments from Oleg Tchernyshyov. We would 
also like to thank Roderich Moessner, John Chalker, and Senthil Todadri for 
critical reading of this manuscript.  Work at the Institute for Quantum Matter 
was supported by the U.S.  Department of Energy, Office of Basic Energy 
Sciences, Division of Material Sciences and Engineering under grant 
DE-FG02-08ER46544.  Access to MACS was provided by the Center for High 
Resolution Neutron Scattering, a partnership between the National Institute of 
Standards and Technology and the National Science Foundation under Agreement 
No. DMR-1508249.  A portion of this research used resources at the Spallation/ 
Neutron Source, a DOE Office of Science User Facility operated by the Oak Ridge 
National Laboratory. This work was supported by the Paul Scherrer Institut by 
providing the supermirror analyzer as a temporary loan to Oak Ridge National 
Laboratory.  We gratefully acknowledge the Johns Hopkins Homewood High 
Performance Cluster (HHPC) and the Maryland Advanced Research Computing Center 
(MARCC), funded by the State of Maryland, for computing resources.

\section*{Author Contributions}
K.~W.~P., A.~S., B.~W., J.~A.~R., and Y.~Q. performed the neutron scattering 
experiments. K.~W.~P. performed the specific heat measurements and analyzed the 
experimental data.  J.~W.~K. and R.~J.~C.  synthesized and characterized the 
single crystal sample. H.~J.~C and S.~Z. performed Monte-Carlo simulations and  
self consistent Gaussian approximation calculations, along with assisting with 
the theoretical interpretation.  K.~W.~P. wrote the manuscript with input from 
all co-authors.  C.~L.~B. oversaw all aspects of the project.  

%
\end{document}